\documentstyle[aps,twocolumn]{revtex}

\begin{document}
\title{Evidence of Collective Charge Behavior in the Insulating State of Ultrathin
Films of Superconducting Metals}
\author{C. Christiansen, L. M. Hernandez, and A. M. Goldman}
\address{School of Physics and Astronomy, University of Minnesota, Minneapolis,\\
MN 55455, USA}
\date{
\today%
}
\maketitle

\begin{abstract}
Nonlinear I-V characteristics have been observed in insulating
quench-condensed films which are locally superconducting. We suggest an
interpretation in terms of the enhancement of conduction by the depinning of
a Cooper pair charge density wave, Cooper pair crystal, or Cooper pair glass
that may characterize the insulating regime of locally superconducting
films. We propose that this is a more likely description than the Coulomb
blockade or charge-anticharge unbinding phenomena.
\end{abstract}

\pacs{PACS numbers: (71.30.+h, 74.25.Dw, 74.76.Db, 74.80.Bj))}

In the context of the Bose-Hubbard model, which is equivalent to the model
of a Josephson junction array with charging, zero-temperature
superconductor-insulator(SI) transitions in two dimensions (2D), tuned by
disorder or magnetic field, are believed to be direct, with metallic
behavior only at the quantum critical point\cite{Fisher}. Recent experiments
have suggested the existence of a significant metallic phase between the
superconductor and insulator. The Stanford group reported this for nominally
superconducting MoGe films in moderate magnetic fields at low temperatures,
and conjectured that it was due to dissipation \cite{Mason}. They later
found ''true'' superconductivity in low fields\cite{Ephron}. Long ago,
metallic behavior was reported in quench-condensed granular films over a
range of thicknesses intermediate between those for which films were
insulating and superconducting\cite{Jaeger}, and it was found more recently
in Josephson junction arrays\cite{Zant}. Re-examination of the theory has
involved the inclusion of aspects of percolation \cite{Percolation},
elaboration of the Bose-Hubbard model\cite{DD,Phillips}, and consideration
of dissipative Bose systems\cite{NGLEE}. Das and Doniach\cite{DD} explained
aspects of Ref. 4 by extending the Bose-Hubbard model to high filling and
including nearest-neighbor as well as on-site Coulomb interactions. Their
phase diagram contains the possibility of an intervening Bose metal phase,
or a direct transition, depending upon the relative magnitudes of the
various Coulomb and Josephson coupling energies. The Bose metal contains
free vortices and antivortices, prevented from Bose condensing by
dissipation. The insulator is a condensate of vortices, or in other
language, a Cooper pair charge density wave (CDW). Phillips and Dalidovich 
\cite{Phillips}, using the original form of the Bose-Hubbard model, also
demonstrated that there was a Bose metallic phase. This followed from a
subtle cancellation in the expression for the conductivity between an
exponentially small population of bosonic quasiparticles and their
associated exponentially long scattering time. The Bose insulator in this
picture results from dissipative processes such as coupling to a heat bath,
or from a high enough level of disorder. Ng and Lee\cite{NGLEE}, using a
different approach to theory of dissipative Bose systems, suggest that a
Bose metal does not exist at zero temperature, but note a crossover regime
at finite temperature that could be identified as a metallic phase.

In this letter we shift the focus to the study of the insulating regime of
granular quench-condensed films. In the past the insulator has not been
carefully studied, as the issue considered was the condition for the onset
of superconductivity. Since local superconductivity has been demonstrated in
such films when they are insulating\cite{Valles}, they should be good
candidates for the application of the Bose-Hubbard model, in which there is
a nonvanishing order parameter amplitude. We have studied Ga films, which
earlier exhibited the most striking metallic regime\cite{Jaeger}. We propose
that the nonlinear I-V characteristics found in the insulating regime may be
evidence of a Bose insulating state, which could be either a Cooper-pair CDW
or Cooper pair crystal. For sufficient disorder it would be a Cooper pair
glass. The nonlinearities appear to be similar to those associated with the
depinning of charge density waves \cite{Gruner}. The behavior of these
clustered films on the insulating side of the SI transition is different
from that of nominally homogeneous films, where the observation of an
orbital magnetoresistance linear in field was interpreted as evidence of
vortices\cite{MarkovicVort}.

Gallium films were grown on glazed alumina substrates held at liquid helium
temperatures. This was done in an ultra high vacuum apparatus in which
cyclic evaporations and {\it in situ} transport measurements\cite{Orr} at
temperatures down to 150 mK could be made. Quoted thicknesses are nominal,
being derived from a calibrated thickness monitor. Standard four-probe
current-biased resistance measurements were made using a low noise dc
current source and a nanovoltmeter. All electrical leads entering the vacuum
chamber were filtered with low-pass interference filters in series with
20-nH\ inductors. This combination exhibited voltage attenuations of 15 dB
at 1 MHz, increasing to 75 dB beyond 100 MHz. Magnetic fields could be
applied up to 12 kG perpendicular and 20 kG parallel to the plane of the
sample. This apparatus is unique in that the evaporants are derived from
commercial Knudsen cells, and the geometry of the cells and the
source-to-substrate distance (60cm) is such that the flux density at the
substrate is uniform to better that one part in $10^{4}.$ This permits the
observation of the dramatic effect of very tiny changes in thickness, which
might be averaged over in film growth systems with less uniformity of flux
density.

To set the stage, we show in Fig. 1 the evolution of $R(T)$ with thickness
for a series of {\it a}-Ga films, adapted from Ref. \cite{Jaeger}. The
thinnest films, shown at the top, exhibit a monotonically increasing $R(T)$
with decreasing temperature. As thickness is increased, $R$ drops near the
bulk $T_{c}$ of {\it a}-Ga, providing evidence for local superconductivity.
At lower temperatures it increases exponentially with decreasing $T$, until
about $2K$, where it levels off. Further increase in thickness results in
metallic behavior at low temperatures. For somewhat thicker films $R(T)$ $%
\sim \exp (\frac{T}{T_{0}})$, which has been interpreted as evidence of
quantum tunneling of vortices\cite{Liu}. At the lowest temperatures for
these films, $R(T)$ is also temperature independent. Further increase in
thickness results in superconductivity. These changes with thickness are
accompanied by changes in the $I-V$characteristics at low temperatures.
Metallic films have linear characteristics, whereas all of the other films
have nonlinear ones. From the onset of local superconductivity to full
superconductivity, nominal film thickness increased from 12.75\AA\ to
14.50\AA .

The leveling off of $R(T)$ of high resistance films was current dependent.
The lower the current, the higher the value of $R(T)$ that would be attained
with decreasing temperature. This is shown in the temperature evolution of
the $I-V$ characteristics in Fig. 2. The differential conductance, $G(V)$,
also shown in Fig. 2, exhibits a threshold voltage, $V_{T}$, which is
independent of temperature, and which decreases with increasing thickness,
both of which are shown in Fig 3.

Heating cannot account for the observed effects for several reasons.
Considering the film to be a thermometer, one can define an ''effective
temperature,'' $T_{eff\text{,}}$ based on $R(T)$. For a range of dissipated
power spanning three orders of magnitude, from $10^{-13}$ to $10^{-10}$ W, $%
T_{eff}$ is well above the thermometer temperature. However it did not
increase with applied power, and in some cases even decreased, implying that
the changes in $R(T)$ are not due to heating and that the film is not acting
as a thermometer. Furthermore, the power at which the resistance leveled off
as films became thicker decreased even when the fractional change in
thickness was small. Actually, the power needed to heat to a particular
temperature should increase rather than decrease with increasing film
thickness, as both heat capacity and thermal conductance become larger. All
of this implies that the nonlinearities are intrinsic. In addition, much
higher resistances were measured in the same apparatus at lower temperatures
for other films not exhibiting either superconductivity or flattening\cite
{MarkovicHop}. Finally, the temperatures at which $R(T)$ became
temperature-independent were well above 1K, making a scenario of not cooling
the electrons unlikely\cite{Clarke}.

Quasiparticle tunneling involving superconductor-insulator-superconductor
(SIS) junctions in the film\cite{Jaeger}can be ruled out because of the
temperature {\it independence} of $V_{T}$ shown in Fig. 3. If $V_{T}$ were a
superconducting gap, it would increase with decreasing temperature. Not
shown is the magnetic field {\it independence} of the threshold, which is
also consistent with it not being attributable to quasiparticle tunneling.
However, the subthreshold conductivity does increase with increasing
temperature and magnetic field suggesting the existence of quasiparticle
transport channels.

A phenomenological picture of charge unbinding has been used to explain
nonlinear $I-V$ characteristics in very much thicker granular Al films,
driven insulating by a magnetic field\cite{WuAdams}. In this picture, a
charge-anticharge pair is formed on neighboring islands by tunneling, and
the members of the pair attract each other. This attraction can be overcome
by an applied electric field which creates free charges. This model, which
includes five free parameters that cannot be independently measured, can be
fit to our data. The fit, which is not shown, does not reproduce the
sharpness of the conductance increase at $V_{T}$.

In Josephson junction arrays with ultra-small junctions \cite{Mooij} and Pd
films \cite{Liu and Price}, nonlinear behavior at high resistances has been
attributed to effects related to the charge Kosterlitz Thouless Berezinskii
(KTB) transition. As a consequence of the electrical duality between the
charge and vortex KTB transitions \cite{Halperin}, nonlinear I-V
characteristics would be expected, with $I\propto V^{a}$, and $a$ exhibiting
a jump from 1 to 3 at the transition temperature. There should also be a
square root cusp in $R(T)$ above the KTB transition temperature. These
effects were not seen. This model is most likely irrelevant because the
screening length in these films is too short for the interaction to be
logarithmic over a substantial distance, a requirement for this scenario\cite
{Keldysh}.

Nonlinearities in the I-V characteristics could be the signature of a
Coulomb blockade\cite{Averin}. The fact that the high-voltage current
asymptote of the I-V characteristic, at low temperatures, extrapolates back
through the threshold voltage $V_{T}$, rather than through the origin, is
consistent with this picture. The macroscopic I-V characteristics of a film
would then be dominated by those associated with tunneling in and out of a
single ''representative island,'' according to the arguments of Ambegaokar,
Langer and Halperin \cite{AHL}. The threshold voltage would be given by the
energy necessary to overcome the Coulomb energy $E_{C}$ in adding charge to
this island, as $V_{T}=E_{C}/e=e/2C$ , where $C$ is the capacitance of the
island. The central argument against this picture is that nonlinearities are
not seen until the temperature corresponding to the onset of local
superconductivity. Values of $V_{T}$ ranging from 6mV to 130mV in films of
different thicknesses, correspond to Coulomb energies much greater than $%
k_{B}T$ at $8.3K,$ where the threshold appears. If the nonlinearities were a
Coulomb blockade effect, they would appear at higher temperatures and would
not require local superconductivity as a precondition for their observation.

We now turn to a picture of the insulating state in which it is CDW-like
configuration of Cooper pairs or a two dimensional Cooper pair crystal. This
idea is mentioned in theoretical papers, but is not treated in detail\cite
{Fisher,DD,Phillips,NGLEE}. We suggest that the nonlinearities are
associated with the depinning of this putative CDW, crystal or glass state.
Features of the depinning that qualitatively resemble the data are the
sharply enhanced conductance at $V_{T}$, and the fact that at low
temperatures, all the I-V curves come together at high bias with the
asymptotic limit of the current extrapolating back to $V_{T}$ \cite{Gruner}.
Atomic force microscope (AFM) pictures of the film sequence discussed in
detail here, which are not shown, were obtained at a thickness of 18.24 \AA
, and after the film was warmed up and removed from vacuum. Although on
warming, annealing of the microstructure takes place, the mesoscale
structure should be largely unchanged, because room temperature, or 20$%
^{0}C, $ is not high enough to result in atomic diffusion over distances
sufficient to change the mesoscale structure. Clusters form during growth,
with the kinetic energy of the impinging atoms making them mobile on the
substrate surface, even though it is attached to a holder cooled to helium
temperature. The AFM pictures showed flat clusters of 200 \AA\ diameter,
which should be representative of the structure at low temperatures. Similar
studies of a thinner film from a different series, revealed smaller
clusters, suggesting a correlation between cluster size and thickness. It
may be possible to consider the film structure of a random array of nearly
identical clusters to be the origin of the pinning of the Cooper pair CDW or
crystal. The decrease of the threshold voltage with film thickness, shown in
the inset of Fig. 3, would then result from a reduction of the areal density
of pinning centers with increasing thickness and physical cluster size.
Scaling of the $I-V$ characteristic near $V_{T}$, which is found in many CDW
systems could not be tested. This was a consequence of our measurements
being current biased. At $V_{T}$ the voltage increased very rapidly,
limiting the data in a manner that made a definitive test of scaling
impossible. What is most important is that the nonlinear behavior manifests
itself only when a local minimum in R(T) is present. Thus the phenomenon
must be associated with Cooper pairing.

The behavior in magnetic field should be noted. In fields up to $12$ $kG$
applied perpendicular and up to $20$ $kG$ applied parallel to the film, $%
V_{T}$ is unchanged, but the onset temperature for local superconductivity
is lowered by up to 1K. Field increases the conductance for $V\prec V_{T}$.
Neither maximum value of field was high enough to completely obliterate the
threshold. The fact that $V_{T}$ is field-independent suggests that the
pinning mechanism is associated with phenomena unaffected by field, i.e.,
defects. The effect of a parallel field on the subgap conductance is smaller
than that of a perpendicular field, but is of the same order of magnitude.
For a $17.2\AA $\ thick Ga film at $4K$, a $12kG$ exhibited a
magnetoconductance of +54\% in a perpendicular field and +20\% in a parallel
field. The parallel field effect cannot be explained by misalignment with
the plane, which was at most $1^{0}.$ These observations suggest that short
range effects of magnetic field on individual grains are comparable to
longer range effects probed by a perpendicular field. Alternatively the
dependence on parallel magnetic field could imply some role for spin degrees
of freedom. The sub-threshold conductance, as mentioned, most likely comes
from quasiparticles. The application of a magnetic field would increase the
number of such quasiparticles participating in transport.

The picture we have presented here as well as the Coulomb blockade and
charge-unbinding effects all involve Coulomb interactions. The difference
between charge-structure depinning and these other phenomena is whether the
behavior is collective, or whether charges individually overcome a Coulomb
energy. Further support for collective behavior is the similarity and even
electrical duality of $G(V)$ characteristics of ''high'' resistance films
with $R(I)$ characteristics of low resistance films. In the latter, the
current threshold is a consequence of the depinning of vortices\cite{Chris}.
The saturation of $G(V)$ and $R(I)$ at low temperatures would imply that
both types of depinning may be quantum phenomena. If collective charge
behavior indeed governs the insulating phase of these 2D superconducting
systems, then it is a member of a larger class of problems including vortex
lattices, charge density waves, Wigner crystals, and magnetic bubbles, in
which it is necessary to understand the dynamics of a periodic structure in
the presence of static substrate disorder. It would then be appropriate to
bring to the problem the ideas used to study such solids\cite{Giam}.

We acknowledge very useful discussions with Sebastian Doniach and Phillip
Phillips, and assistance with AFM from Pete Eames. This work was supported
by the National Science Foundation under grant \# NSF/DMR- 987681.



\begin{figure}[tbp]
\caption{Evolution of the temperature dependence of the sheet resistance for
a series of Ga films, adapted from Ref. 4. Film thicknesses range from 12.75
\AA\ to 26.67 \AA\ and increase from top to bottom. Thickness was changed in
increments of 0.05\AA . Some data has been omitted for clarity.}
\label{fig1}
\end{figure}
\begin{figure}[tbp]
\caption{(a) The evolution of current as a function of voltage with
temperature in a 18.2 \AA\ Ga film. The temperatures are 8, 7, 6, 5, 4, 3,
2, 1.5, 1, 0.5, 0.15 K from top to bottom, with curves overlappping at 2K
and below. (b) Differential conductance, dI/dV, as a function of voltage for
the temperatures on the same film. A temperature independent threshold
voltage of about 6 mV can be seen where the conductance suddenly increases.}
\label{fig2}
\end{figure}

\begin{figure}[tbp]
\caption{Threshold voltage as a function of temperature of an 18.2 \AA\ Ga
film. Inset: threshold voltage as a function of thickness for four Ga films
in a series.}
\label{fig3}
\end{figure}



\begin{references}
\bibitem{Fisher}  M. P. A. Fisher, G. Grinstein, and S. M. Girvin, Phys.
Rev. Lett. {\bf 64}, 587 (1990); Min-Chul Cha {\it et al., }Phys. Rev. B 
{\bf 44}, 6883 (1991).

\bibitem{Mason}  D. Ephron, A. Yazdani, A. Kapitulnik and M. R. Beasley,
Phys. Rev. Lett. {\bf 76}, 1529 (1996); N. Mason and A. Kapitulnik, Phys.
Rev. Lett. {\bf 82}, 5341 (1999).

\bibitem{Ephron}  N. Mason and A. Kapitulnik, cond-mat/0006138 (unpublished).

\bibitem{Jaeger}  H. M. Jaeger, D. B. Haviland, B. G. Orr and A. M. Goldman,
Phys. Rev. B {\bf 34}, 4920 (1986) and Phys. Rev. B {\bf 40}, 182 (1989).

\bibitem{Zant}  H. S. J. van der Zant, {\it et al}., Phys. Rev. B {\bf 54},
10081 (1996).

\bibitem{Percolation}  E. Shimshoni, A. Auerbach and A. Kapitulnik, Phys.
Rev. Lett. {\bf 80}, 3352 (1998); A. Kapitulnik, N. Mason, S. Kivelson, S.
Chakravarty, Phys. Rev. B, {\bf 63}, 125322 (2001).

\bibitem{DD}  D. Das and S. Doniach, Phys. Rev. B. {\bf 60}, 1261(1999).

\bibitem{Phillips}  D. Dalidovich and P. Phillips, cond-mat/0005119
(unpublished).

\bibitem{NGLEE}  Tai Kai Ng and Derek K. K. Lee, Phys. Rev. B 63, 144509
(2001).

\bibitem{Valles}  J. M. Valles Jr., S.-Y. Hsu, R. C. Dynes, J. P. Garno,
Physica B, {\bf 197}, 522 (1994).

\bibitem{Gruner}  G. Gr\"{u}ner, {\it Density Waves in Solids},
(Addison-Wesley, Reading, Massachusetts, 1994) and references therein.

\bibitem{MarkovicVort}  N. Markovic {\it et al}, Phys. Rev. Lett., {\bf 81},
701 (1998).

\bibitem{Orr}  B. G. Orr and A. M. Goldman, Rev. Sci. Instr. {\bf 56}, 1288
(1985).

\bibitem{Liu}  Y. Liu, D. B. Haviland, L. I. Glazman, and A. M. Goldman,
Phys. Rev. Lett. {\bf 68}, 2224 (1992).

\bibitem{MarkovicHop}  N. Markovi\'{c}, C. Christiansen, D.E. Grupp, A.M.
Mack, G. Martinez-Arizala, and A.M. Goldman, Phys. Rev. B {\bf 62}, 2195
(2000).

\bibitem{Clarke}  F.C. Wellstood, C. Urbina, and J. Clarke, Phys. Rev. B 
{\bf 49}, 5942 (1994).

\bibitem{WuAdams}  W. Wu and P. W. Adams, Phys. Rev. B {\bf 50}, 13065
(1994).

\bibitem{Mooij}  J. E. Mooij {\it et. al.}, Phys. Rev. Lett. {\bf 65}, 645
(1990).

\bibitem{Liu and Price}  Y. Liu and J. C. Price, Physica B {\bf 194-196},
1351 (1994).

\bibitem{Halperin}  B. I. Halperin and D. R. Nelson, J. Low. Temp. Phys. 
{\bf 36}, 1165 (1979).

\bibitem{Keldysh}  L. V. Keldysh, Pis'ma Zh. Eksp.Teor. Phys. {\bf 29}, 716
(1979) [JETP Lett. {\bf 29}, 659 (1979)].

\bibitem{Averin}  D. A. Averin and K. K. Likharev, in {\it Mesoscopic
Phenomena in Solids}, ed. B. L. Altshuler, P. A. Lee and R. A. Webb, p. 173
(North-Holland, Amsterdam 1991).

\bibitem{AHL}  V. Ambegaokar, B. I. Halperin and J. S. Langer, Phys. Rev. B%
{\bf 4}, 2612 (1971).

\bibitem{Chris}  C. Christiansen, Doctoral Dissertation, University of
Minnesota, 2001, unpublished.

\bibitem{Giam}  Pierre Le Doussal and Thierry Giamarchi, Phys. Rev. B {\bf 57%
}, 11336 (1998) and references cited therein.
\end{references}
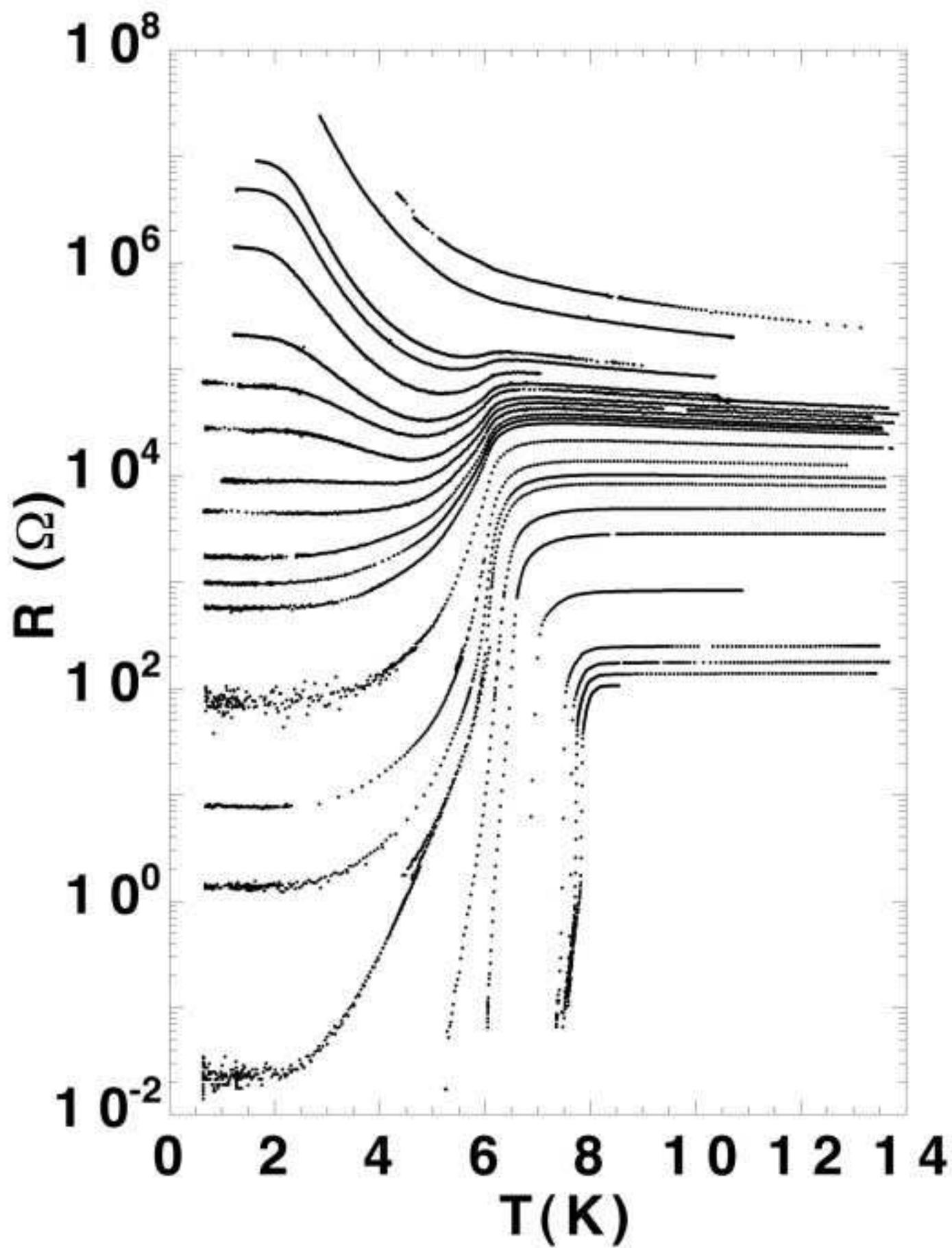
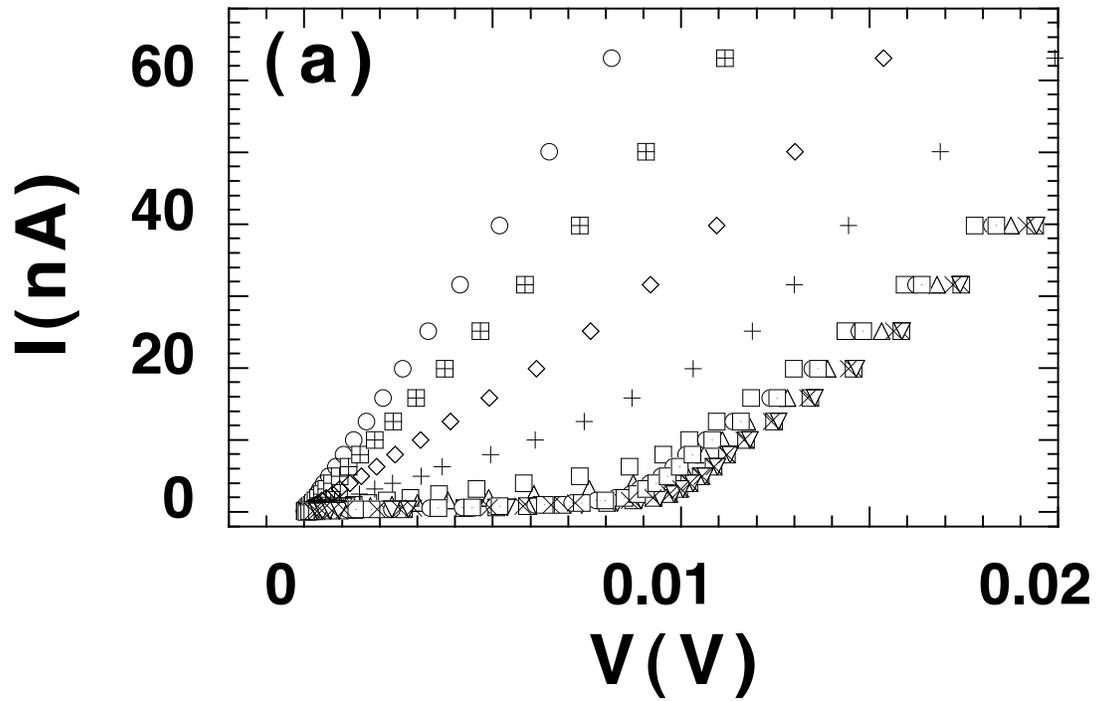
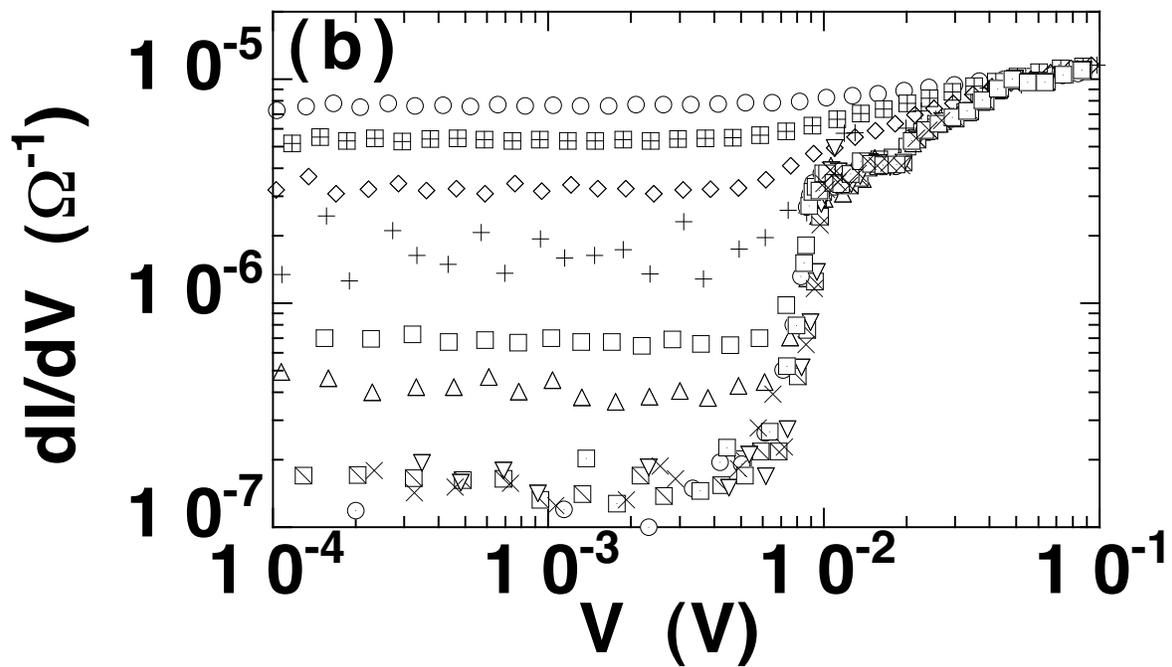
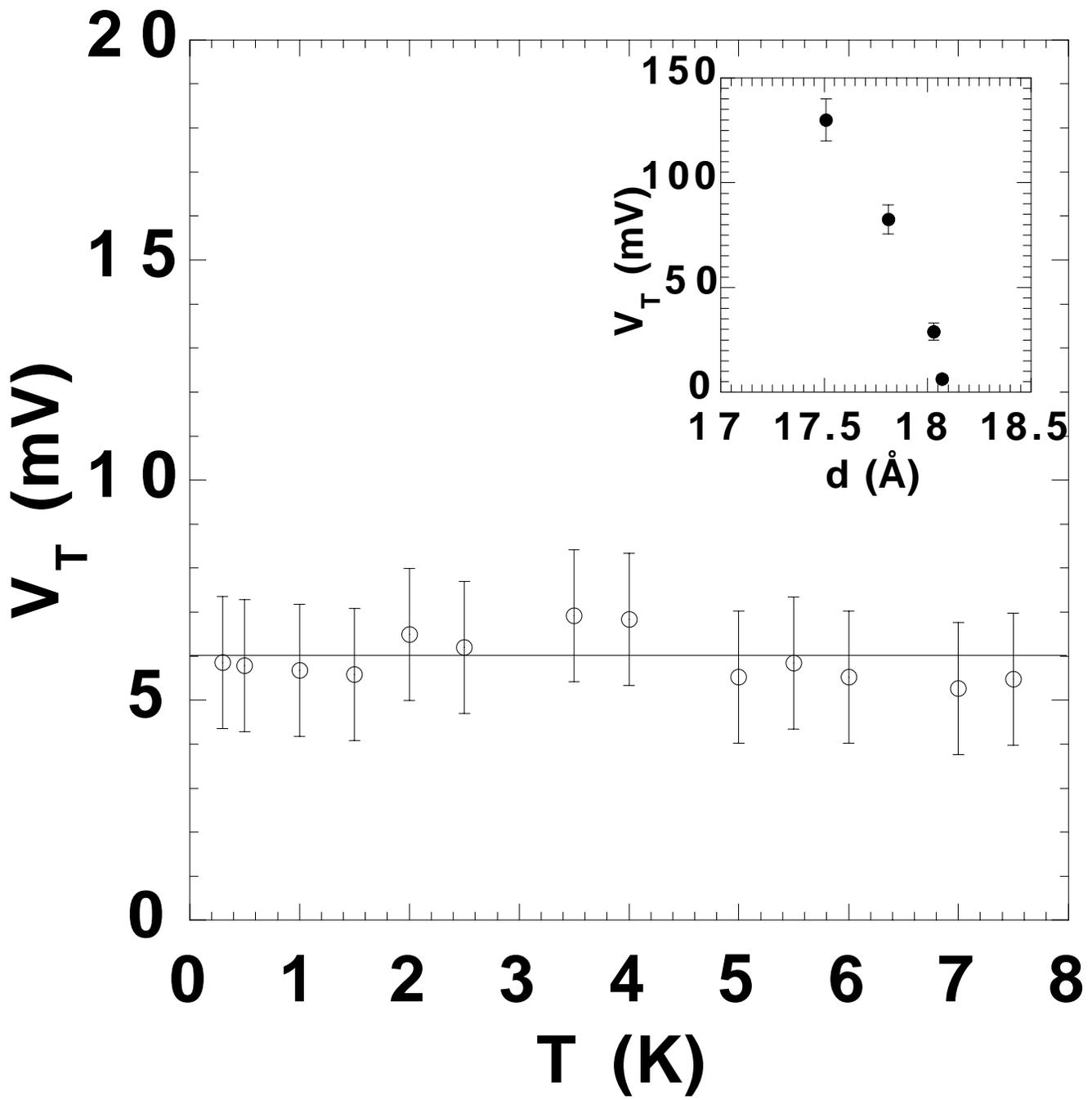
\end{document}